\documentclass[superscriptaddress,twocolumn,showpacs,preprintnumbers,amsmath,amssymb,floatfix,nofootinbib]{revtex4-2}

\usepackage{color}   %option added by cyw eg \pagecolor{blue} \color{red}
\usepackage{graphicx}% Include figure files
\usepackage{dcolumn}% Align table columns on decimal point

\begin{document}

\def\bb    #1{\hbox{\boldmath${#1}$}}

\title{Sub-Poissonian multiplicity distributions in jets produced in hadron collisions}

\author{Han Wei Ang}
\email{ang.h.w@u.nus.edu}
\affiliation{Department of Physics, National University of Singapore, Singapore 117551}
\author{Maciej Rybczy\'nski}
\email{maciej.rybczynski@ujk.edu.pl}
\affiliation{Institute of Physics, Jan Kochanowski University, 25-406 Kielce, Poland}
\author{Grzegorz Wilk}
\email{grzegorz.wilk@ncbj.gov.pl}
\affiliation{ National Centre for Nuclear Research, Warsaw 00-681, Poland}
\author{Zbigniew W\l odarczyk}
\email{zbigniew.wlodarczyk@ujk.edu.pl}
\affiliation{Institute of Physics, Jan Kochanowski University, 25-406 Kielce, Poland}

\begin{abstract}
In this work we show that the proper analysis and interpretation of the experimental data on the multiplicity distributions of charged particles produced in jets measured in the ATLAS experiment at the LHC indicates their sub-Poissonian nature. We also show how, by using the recurrent relations and  combinants of these distributions, one can obtain new information contained in them and otherwise unavailable, which may broaden our knowledge of the particle production mechanism.
\end{abstract}


\maketitle
\section{Introduction}
\label{Introduction}

The experimentally measured multiplicity distributions $P (N)$ of the produced particles are the main source of information about the dynamics of their production processes \cite{Kittel,Book-BP}. In the theoretical description they are characterized by the generating functions,
\begin{equation}
G(z) = \sum_{N=0}^{\infty} P(N) z^N, \label{GF}
\end{equation}
such that
\begin{equation}
P(N) = \frac{1}{N!}\frac{d^N G(z)}{d z^N}\bigg|_{z=0}. \label{PN}
\end{equation}
Preliminary information about $P (N)$ is provided by the moments of this distribution,
\begin{equation}
m_k = \sum_{N=0}^{\infty} (N - c)^k P(N). \label{Mk}
\end{equation}
In many cases, it is sufficient to analyze only the two lowest moments, namely the mean value $\langle N\rangle = m_1$ being the first raw moment $(c = 0)$, and variance $Var(N) = m_2$, being the second central moment ($c= \langle N\rangle$).

Another way to characterize $P(N)$ is through a recursive formula,
\begin{equation}
(N + 1)P(N + 1) = g(N) P(N), \label{Method2}
\end{equation}
that connects adjacent values of $P(N)$ for the production of $N$ and $(N + 1)$ particles. It is assumed here that every $P(N)$  is determined only by the next lower $P(N - 1)$ value. In other words, a relationship with others $P(N - j )$ for $j > 1$ is indirect. The final algebraic form of $P(N)$ is determined by the function $g(N)$. In its simplest form, $g(N)$ is assumed to be a linear function of $N$ given by $g(N) = \alpha + \beta N$. This form is enough to define commonly known and widely used distributions like Poisson Distribution (PD), for which $\beta = 0$), Binomial Distribution (BD) (for which $\beta < 0$)  or Negative Binomial Distribution (for which $\beta > 0$). In general, by selecting the appropriate model, the form $g(N)$ can be chosen in such a way that the corresponding $P(N)$ describes the experimental data (for example, by introducing higher order terms \cite{HC} or by using its more involved forms \cite{ChK,Zg}).

The more promising approach is to use $ g(N) $ which contains information about the interrelationship between the multiplicity $N$ and all smaller multiplicities recursively,
\begin{equation}
(N + 1)P(N + 1) = \langle N\rangle \sum^{N}_{j=0} C_j P(N - j). \label{rr}
\end{equation}
The memory of that relationship is encoded in coefficients $C_j$ called {\it modified combinants}. They were introduced and intensively discussed in \cite{ST,Ours1,IJMPA, Ours2,Ours3,Ours-EPJA,IJMPA-WW,Zborovsky}. By inverting the recursion (\ref{rr}) we obtain an equation that allows us to determine $C_j$ from the measured $P(N)$,
 \begin{equation}
\langle N\rangle C_j = (j+1)\left[ \frac{P(j+1)}{P(0)} \right] - \langle N\rangle \sum^{j-1}_{i=0}C_i \left[ \frac{P(j-i)}{P(0)} \right] \label{rCj}
\end{equation}
(provided we have sufficient statistics). Modified combinants are closely related to {\it combinants}  $C^{\star}_j$,
\begin{equation}
C_j = \frac{j+1}{\langle N\rangle} C^{\star}_{j+1}, \label{CstarC}
\end{equation}
introduced in \cite{Combinants-1,Combinants-2,VVP} by means of the generating functions $G(z)$ as
\begin{equation}
\langle N\rangle C^{\star}_j = \frac{1}{j!} \frac{ d^{j+1} \ln G(z)}{d z^{j+1}}\bigg|_{z=0}, \label{GF_Cj}
\end{equation}
which have been discussed and used in many publications \cite{Kittel, PCarruthers, Book-BP, CombUse1a, CombUse3, CombUse4, CombUse5}.

Modified combinants are complementary to the commonly used factorial moments, $F_q$ and cumulant factorial moments, $K_q$ (see Appendix \ref{FqKq} for details). They differ in that while $C_j$'s depend only on multiplicities smaller than their rank, $K_q$'s require the knowledge of all $P(N)$'s and are therefore very sensitive to possible limitations of the available phase space \cite{Kittel,Book-BP}. However, both $C_j$ and $K_q$ share the property of additivity. It turns out that most of the measured multiplicity distributions $P(N)$ give oscillatory combinants $C_j$ with increasing index $j$ which arise due to the presence of a BD component in the measured $P(N)$ \cite{Ours1,IJMPA, Ours2,Ours3,Ours-EPJA,IJMPA-WW}.

Combinants are believed to be best suited for the study of sparsely populated areas of phase space (while cumulants are better suited for the study of densely populated areas) \cite{Kittel, Book-BP}. This feature makes them a potentially important tool for studying $P(N)$ in jets where the number of produced particles is small (on the order of $\sim 10$). However, the ATLAS data \cite{AtlasJ}  do not include $P(0)$, which is crucial for their determination. This means that in our analysis, we must extrapolate the recurrent relation in Eq. (\ref{Method2}) to evaluate $P(0)$ and use the combinants only for additional verification of our conclusions.

In the next Section we provide details concerning ATLAS data with particular attention to the fact that the measured multiplicity distributions in the jets are clearly sub-Poissonian in character (with details depending on the phase-space covered). This observation will be our main point for further discussion and calculations described in Sections \ref{PossibleExpl} and \ref{sum-res}.  Section \ref{Concl} summarizes and concludes our work.

\section{Multiplicity distributions of particles in ATLAS jets}
\label{P(N)inJ}

ATLAS data \cite{AtlasJ} of jets measured in proton-proton collisions at an energy $\sqrt{s}=7$ TeV (using a minimum bias trigger) were taken (data used here come from \cite{AtlasJD}). Jets were reconstructed using the anti-$k_t$ algorithm applied to charged particles produced in very narrow cones defined by radius parameter $R=\sqrt{\Delta \eta^2 + \Delta \phi^2}$ (where $\Delta \phi$ and $\Delta \eta$ are the azimuthal angle and the pseudorapidity of the hadrons relative to that of the jet respectively. Here, $\eta =- \ln \tan \theta$, with $\theta$ being the polar angle), with $R=0.4$ and again at $R = 0.6$. Only data with high statistics within acceptable regions of phase space are selected. They are obtained over $5$ different transverse momentum ranges across $4$ rapidity ranges, giving a total of $20$ possible combinations for each of the two radius parameters (with radius parameters $R=0.4$ and $R=0.6$, respectively).

So far, the ATLAS data have been carefully analyzed in terms of possible self-similarity between $p_T$ distribution of jets and $p_T$ distributions of particles in these jets \cite{WWJets}. Indeed, their detailed analysis clearly indicates the self-similarity of the particle distributions in jets and the distributions of the jets themselves, indicative of the existence of a common mechanism behind all these processes.

\begin{figure}[t]
\begin{center}
\includegraphics[scale=0.48]{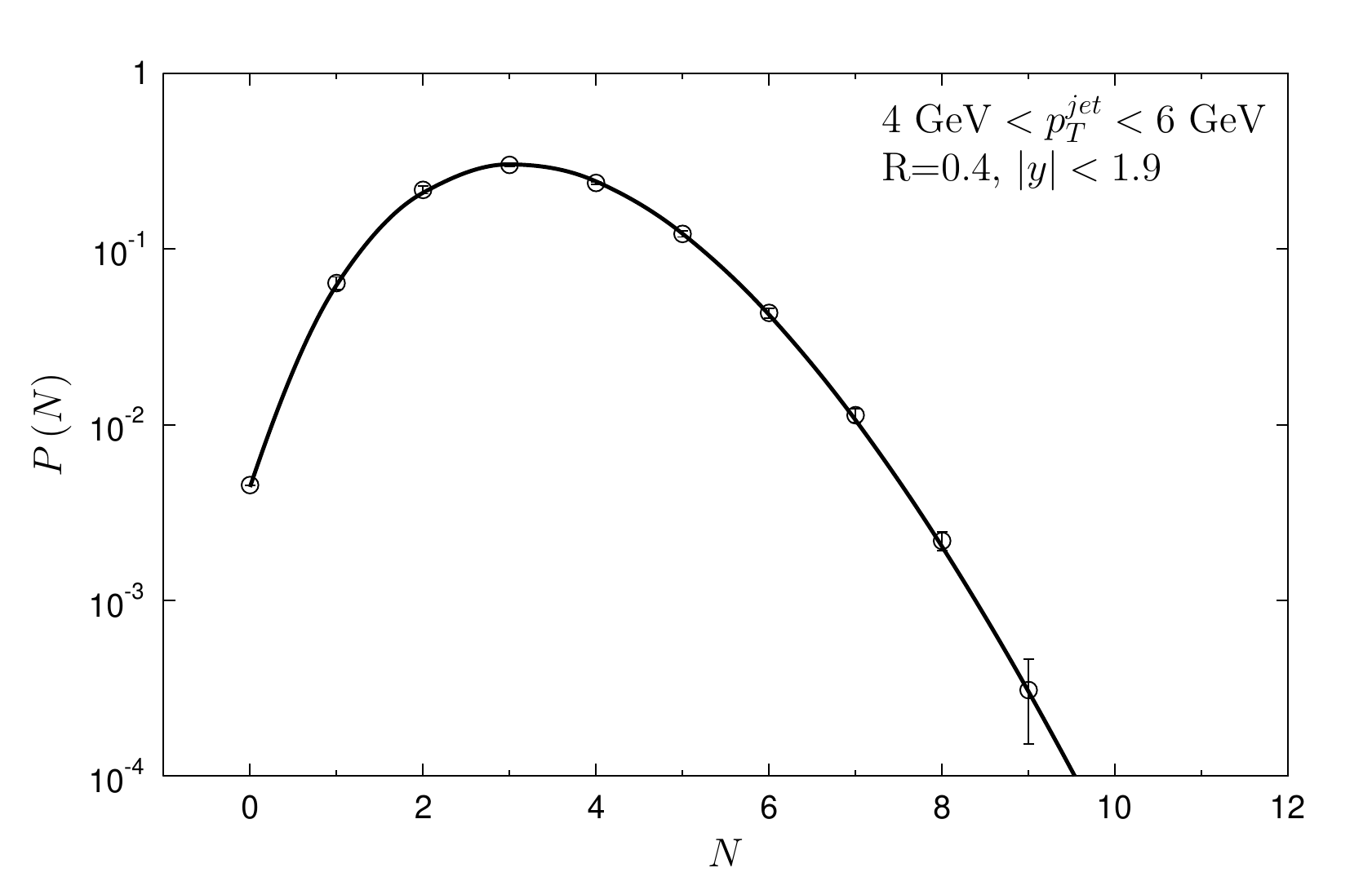}
\end{center}
\vspace{-3mm}
\caption{Multiplicity distribution of charged particles per jet with $4~GeV < p_T^{jet} < 6~GeV$, over the full measured rapidity range $|y|<1.9$, with radius parameter $R=0.4$. Points show data from ATLAS experiment \cite{AtlasJ}. $P\left(N=0\right)$ comes from extrapolation of experimental recurrent relation $g\left(N\right)$ to $N=0$. The curve fitting this data comes from Eq. (\ref{ZW-5}) with parameters  $\alpha =14.1$  and $\delta=1.07$.}
\label{SPD-1}
\end{figure}

\begin{figure}[b]
\begin{center}
\includegraphics[scale=0.48]{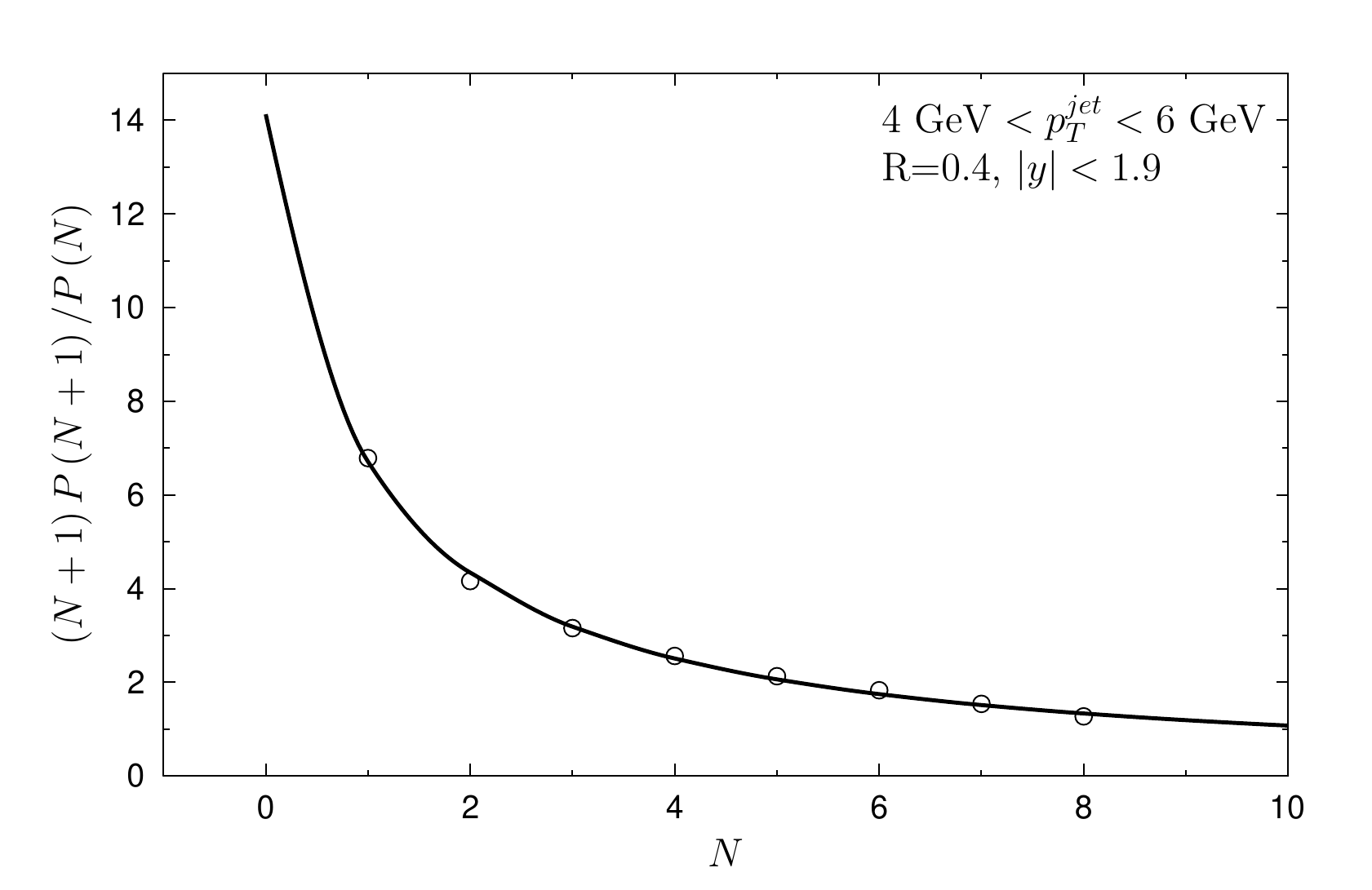}
\end{center}
\vspace{-3mm}
\caption{Recurrent relations. Points show $g(N)$ for experimental $P(N)$ from Fig.~\ref{SPD-1}. Curve fit using $g(N)$ from Eq.~(\ref{ZW-1}) with $\alpha =14.1$  and $\delta = 1.07$.}
\label{SPD-3}
\end{figure}

Taking advantage of the fact that ATLAS also publishes data for multiplicity distributions of particles produced in observed jets, we extend this analysis to study the nature of these distributions. The first observation is that the multiplicity distributions of the particles in the jets observed in the ATLAS experiment are sub-Poissonian, cf. Fig. \ref{SPD-1} where $Var(N) < \langle N\rangle$. Though BD is a possible sub-Poissonian distribution for $g(N)$, the plot in Fig. \ref{SPD-3} derived from the recursive relationship in Eq. (\ref{Method2}) is non-linear.

It means therefore that one should switch to scenario inspired by the possible non-linear form of recursion $g(N)$ such as
\begin{equation}
g(N) = (N+1)\frac{P(N+1)}{P(N)} = \frac{\alpha}{(N+1)^{\delta}}. \label{ZW-1}
\end{equation}
As shown in Fig. \ref{SPD-3}, such form with $\delta = 1.07$ fits data very well. Therefore, we note that for
\begin{equation}
(N+1)\frac{P(N+1)}{P(N)}=\alpha \label{ZW-2}
\end{equation}
we have the PD with,
\begin{equation}
P(N) = c\frac{\alpha^N}{N!}\quad {\rm where}\quad \alpha = \langle N\rangle,\quad c=\exp(-\alpha).\label{ZW-3}
\end{equation}
If we change this recursive relationship to a nonlinear one given by
\begin{equation}
(N+1)^{1+\delta}\frac{P(N+1)}{P(N)}=\alpha, \label{ZW-4}
\end{equation}
we get sub-Poissonian distribution,
\begin{equation}
P(N) = c\frac{\alpha^N}{(N!)^{1+\delta}},\label{ZW-5}
\end{equation}
where $c=P(0)$ is a normalization factor. In Fig. \ref{SPD-1}  we present a comparison of this multiplicity distribution with the experimental data from ATLAS.

\begin{figure}[t]
\begin{center}
\includegraphics[scale=0.48]{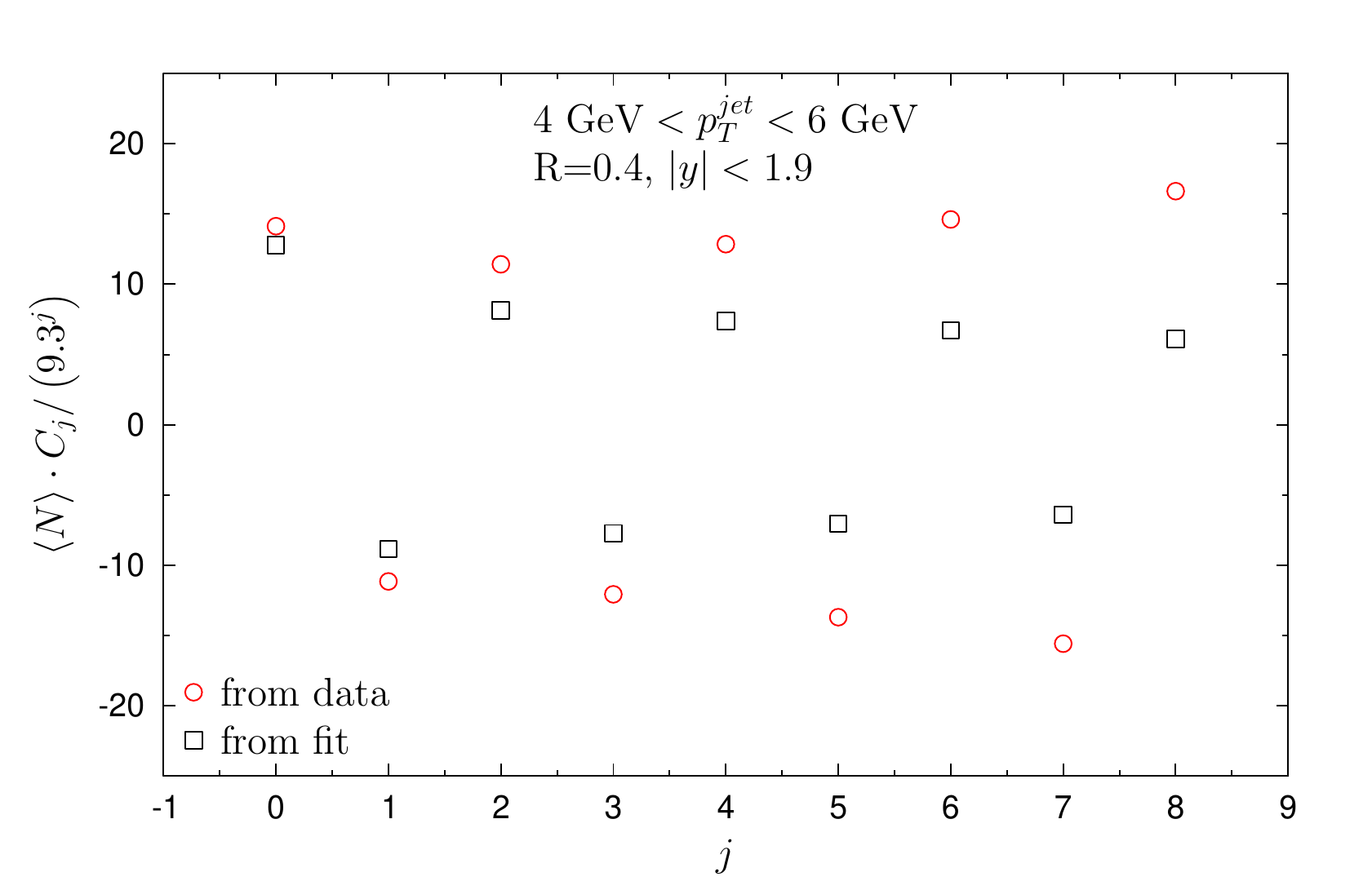}
\end{center}
\vspace{-3mm}
\caption{Comparison of $C_j$ for multiplicity distributions from Fig. \ref{SPD-1}. Red circles: $C_j$ from data on $P(N)$ from Fig. \ref{SPD-1}. Black squares: from theoretical $P(N)$ defined by Eq.(\ref{ZW-5}).}
\label{SPD-2}
\end{figure}

Once we know $P(0)$ from extrapolation of experimental recurrent relation $g\left(N\right)$ to $N=0$, we are able to determine the corresponding modified combinants from the measured multiplicity distributions $P(N)$ using Eq. (\ref{rCj}) (which is important because in ATLAS, the $P(0)$ data points are not directly measured). The $C_j$'s derived in this way are plotted in Fig. \ref{SPD-2} as red circles. They can now be compared  with the combinants obtained in the same way but from the theoretical sub-Poissonian distribution $P(N)$ given by Eq. (\ref{ZW-5}). The corresponding results are plotted in  Fig. \ref{SPD-2} as black squares. We observe characteristic oscillatory behaviour of modified combinants with only rough amplitude agreement. Despite nice agreement of multiplicity distributions, shown in Fig.~\ref{SPD-1}, the mentioned combinants indicate difference between fit and experimental data.

In the event where  $\delta =1$ Eq. (\ref{ZW-5}) can be written in a closed form given by,
\begin{equation}
P(N) = \frac{1}{I_0(2\sqrt{\alpha})}\frac{\alpha^N}{(N!)^2}, \label{NZW-1}
\end{equation}
where
\begin{equation}
\alpha = \langle N^2\rangle\quad {\rm and}\quad c=P(0)=\frac{1}{I_0(2\sqrt{\alpha})},\label{ZW-6}
\end{equation}
with $I_0$ being the Bessel function of the first kind. In this case ($\delta=1$) the theoretical $C_j$'s are given by the recurrent relation,
\begin{equation}
\langle N\rangle C_j = \frac{(j+1)\alpha^{j+1}}{[(j+1)!]^2} - \langle N\rangle \sum_{i=0}^{j-i}C_i\frac{\alpha^{j-i}}{[(j-i)!]^2}, \label{ZW-7}
\end{equation}
and are of the form
\begin{equation}
\langle N\rangle C_j = (-1)^j \beta_j \alpha^{j+1} \label{ZW-8}
\end{equation}
where the numbers $\beta_j$ are rational: $\beta_0=1$, $\beta_1=\frac{1}{2}$, $\beta_2=\frac{1}{3}$, $\beta_3 = \frac{11}{48}$, $\beta_4 = \frac{19}{120}$, $\beta_5 = \frac{473}{4320}, \dots $. They were first calculated by Euler in relation to the positive zeros $\gamma_l$ of the Bessel function $J_0(z)$ as \cite{BesselF}
\begin{equation}
\beta_{j+1} = \sum_{l=1}^{\infty} \left(\frac{2}{\gamma_l}\right)^{2(j+l)},\qquad j=0,1,2,\dots .\label{ZW-9}
\end{equation}
For $j\ge 1$ coefficients $\beta_j$ can be approximated by
\begin{equation}
\beta_j \cong \exp\left[ - \frac{j+1}{e}\right]. \label{ZW-10}
\end{equation}

\begin{figure}[b]
\begin{center}
\includegraphics[scale=0.48]{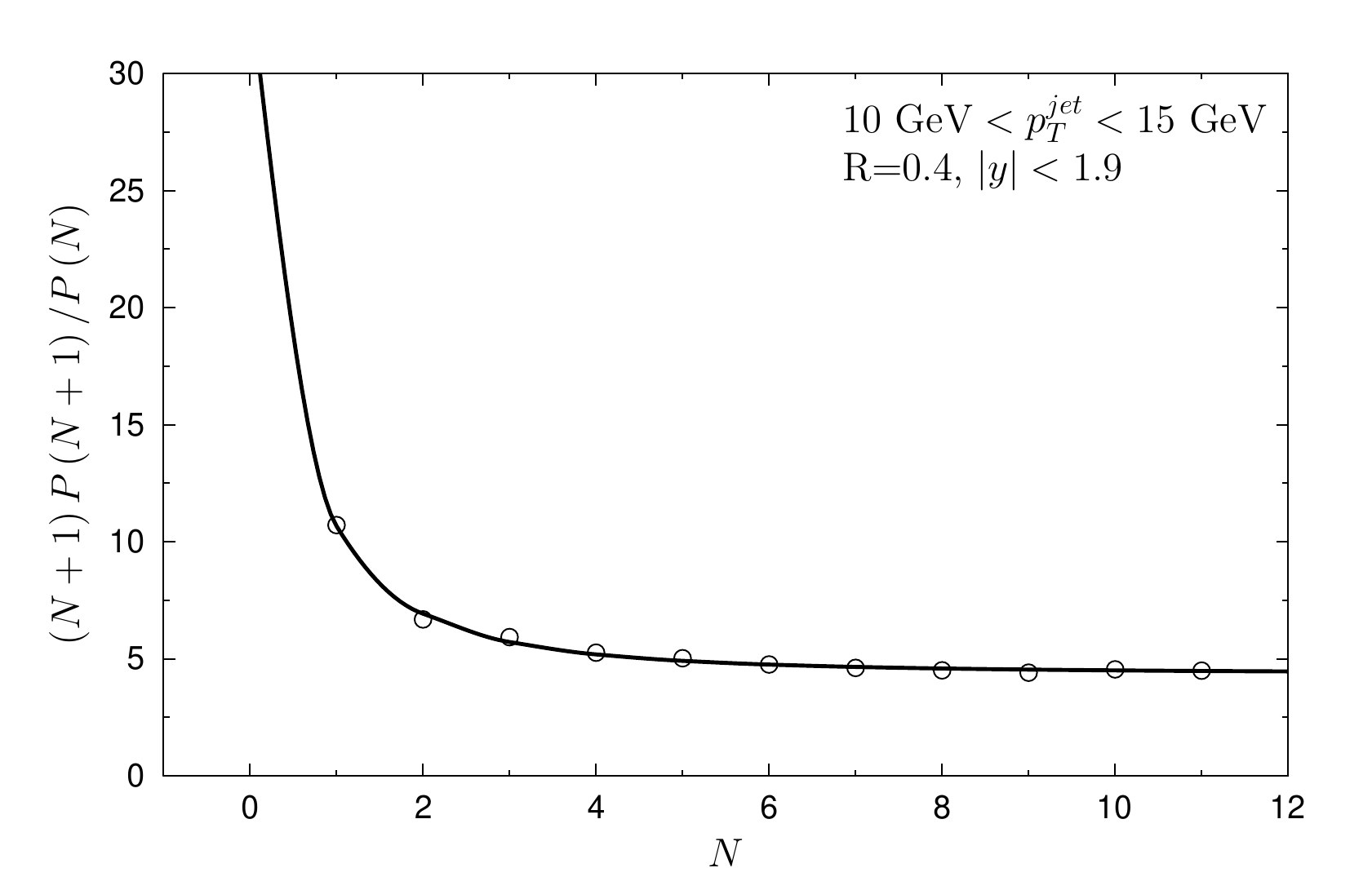}
\end{center}
\vspace{-3mm}
\caption{Recurrent relations $g(N)$ for multiplicity distributions $P(N)$ over the full measured rapidity range $|y|<1.9$, in jets with radius parameter $R=0.4$ and transverse momentum range $10~ GeV < p_T^{jet} < 15~ GeV$. Points: $g(N)$ from experimental data \cite{AtlasJ}. Curve: fit using $g(N)$ from Eq. (\ref{NZW-2}) with parameters: $\alpha = 29.4$, $\delta = 2.20$ and $\alpha_0 = 4.37$.}
\label{SPD-4}
\end{figure}

\begin{figure}
\begin{center}
\includegraphics[scale=0.48]{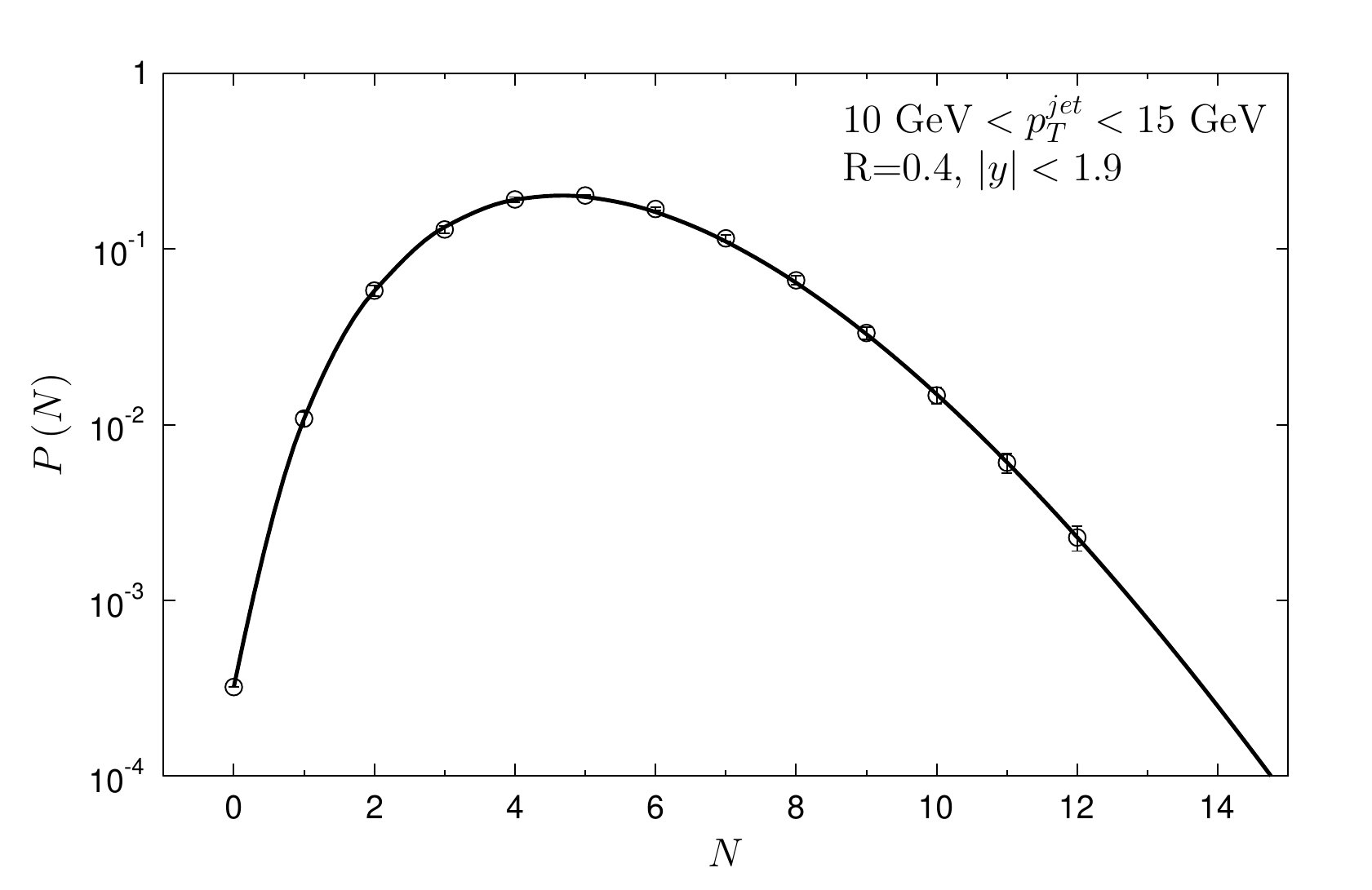}
\end{center}
\vspace{-3mm}
\caption{Points: $P(N)$ from ATLAS data for jets with $10~ GeV < p_T^{jet} < 15~GeV$ and radius parameter $R=04$, over the full measured rapidity range $|y|<1.9$. The curve fitting these data comes from Eq.(\ref{NZW-3}) with parameters: $c = 3.2\cdot 10^{-4}$, $\alpha = 29.4$, $\alpha_0 = 4.37$ and $\delta = 2.20$.}
\label{SPD-5}
\end{figure}
\vspace{-3mm}
\begin{figure}[b]
\begin{center}
\includegraphics[scale=0.48]{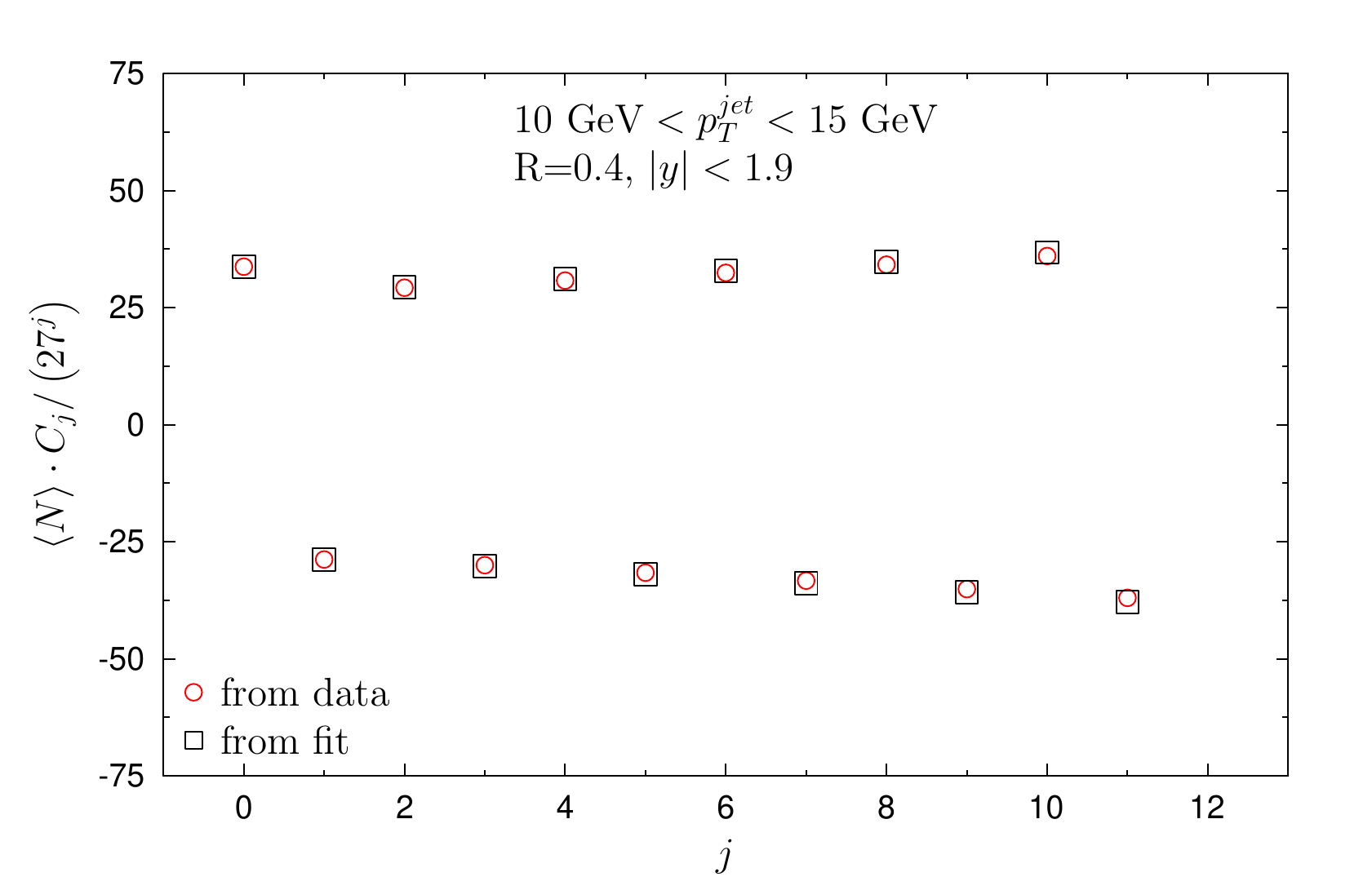}
\end{center}
\vspace{-3mm}
\caption{Comparison of $C_j$ for multiplicity distribution in jets. Red circles: $C_j$ from data on $P(N)$, black squares: from theoretical $P(N)$ defined by Eq.(\ref{NZW-3}). Parameters the same as in Fig. \ref{SPD-5}.}
\label{SPD-6}
\end{figure}

At this point, it is worth noting that $C_j$ can help in the search for the correct $P (N)$. Based on our experience thus far, let us assume, that strongly oscillating $C_j$ (especially with period, as seen in Fig. \ref{SPD-2}) indicate the presence of a single-component BD in some form. On closer inspection, it is clear that this cannot be the case since the amplitude of oscillations of $C_j$ in Fig. \ref{SPD-2} grows approximately as $9.3^j$ with $j$. Should these $C_j$ originate from a single-component BD with amplitudes given by $[p/(1-p)]^j$, it would mean that $p > 0.8$. In addition, to reproduce $\langle N\rangle = Kp \sim 3.3$ as observed in data, one would require $K < 5$. Taken together, these would limit us to multiplicities of $N<5$. This is in contradiction with the measured $P(N)$ where the observed multiplicities $N=5$. By continuing to stick to BD, our previous experiences \cite{Ours1, Ours2, Ours3, Ours-EPJA} tell us that a potential solution might be to use the sum of two BD's instead of one. However, as we will show in the Appendix \ref{2BD}, it is not possible to describe both $P(N)$ and its corresponding $C_j$ with this approach.

Continuing the approach based on $g(N)$, it turns out that with the increase in $p_T$ of jets (corresponding to an increase of $\langle N\rangle$ in our case), we observe a deviation from the form of $g(N)$ given in Eq. (\ref{ZW-1}). The modified $g(N)$ can be made to describe data if expressed as a recursive relation given by (see Fig. \ref{SPD-4})
\begin{equation}
g(N) = (N+1)\frac{P(N+1)}{P(N)} = \frac{\alpha}{(N+1)^{\delta}} + \alpha_0, \label{NZW-2}
\end{equation}
leading to the multiplicity distribution (see Fig. \ref{SPD-5})
\begin{equation}
P(N) = \frac{c}{N!}\prod_{i=1}^N\left( \frac{\alpha}{i^{\delta}} + \alpha_0\right) \label{NZW-3}
\end{equation}
with $c=P(0)$. The corresponding $C_j$ are shown in Fig. \ref{SPD-6}.

Note that for integer parameter $\delta$ , Eq. (\ref{NZW-3})) has closed analytical form, namely for $\delta = 1$
\begin{equation}
P(N) = P(0) \frac{\alpha_0^N}{(N!)^2} \frac{\Gamma\left( 1 + \frac{\alpha}{\alpha_0} + N \right)}{\Gamma\left( 1 + \frac{\alpha}{\alpha_0} \right)}, \label{AdZW-1}
\end{equation}
while for $\delta = 2$
\begin{eqnarray}
P(N) &=& P(0) \frac{\alpha_0^N}{(N!)^3}\cdot\nonumber\\
 &&\cdot \frac{\Gamma\left(1 - \sqrt{\frac{-\alpha}{\alpha_0}} + N\right)}{\Gamma\left(1 - \sqrt{\frac{-\alpha}{\alpha_0}}\right)}
 \frac{\Gamma\left(1 + \sqrt{\frac{-\alpha}{\alpha_0}} + N\right)}{\Gamma\left(1 + \sqrt{\frac{-\alpha}{\alpha_0}}\right)}, \label{AdZW-2}
\end{eqnarray}
and for higher value of $\delta$ we have product of $\delta$ Pochhammer symbols,
\begin{equation}
\left( x_l\right)_N = \frac{\Gamma\left( x_i + N\right)}{\Gamma\left( x_i\right)}, \label{AdZW-3}
\end{equation}
where $x_l = 1 + (-1)^l (-1)^{l/\delta}\left( \alpha/\alpha_0\right)^{1/\delta}$ for $l = 0, 1, \dots, \delta -1$.

\section{Possible explanation: Multiplicity dependent birth and death rates}
\label{PossibleExpl}

To interpret the results shown in the previous section, note that Eq. (\ref{ZW-5}) actually represents the so-called COM-Poisson distribution introduced by Conway and Maxwell \cite{COM1962} as a model for steady state queuing systems with state-dependent arrival or service rates (in other words, birth-death process with Poisson arrival rate and exponential service rate). It was rediscovered in \cite{Shmuel} where the term {\it Conway-Maxwell-Poisson} was proposed and a detailed study of its properties and applications was performed. More recent studies can be found in \cite{ChI, ChShmuel}. To our best knowledge, this distribution has not been used in the analysis of multiplicity distributions of particles produced in multiparticle production processes.

We will now show that the form of the COM-Poissonian distribution  can be obtained from a stochastic Markov process with multiplicity-dependent birth and death rates denoted by $\lambda_N$ and $\mu_N$, respectively \cite{LZH}. Let  $P(N,t)$ be the probability of having $N$ particles at time $t$ and let us consider a very general birth-death process given by the following equations:
\begin{eqnarray}
\!\!\!\!\!P'(0,t) &=& -\lambda_0 P(0,t) + \mu_1 P(1,t), \label{NZW-4a}\\
\!\!\!\!\!P'(N,t) &=& -\left( \lambda_N + \mu_N \right) P(N,t) + \nonumber\\
&&+ \lambda_{N-1} P(N-1,r) + \mu_{N+1} P(N+1,t). \label{NZW-4b}
\end{eqnarray}
If we assume the forms
\begin{equation}
\lambda_N = \frac{\lambda}{(N+1)^a}\quad{\rm and}\quad \mu_N=N^b \mu, \label{NZW-5}
\end{equation}
we get
\begin{eqnarray}
&& -\left[ \frac{\lambda}{(N+1)^a} + N^b \mu \right]P(N) + \nonumber\\
&&\hspace{8mm} + \frac{\lambda}{N^a}P(N-1) + (N+1)^b \mu P(N+1) = 0\label{NZW-6}
\end{eqnarray}
for the steady state, where $P'(N,t)=0$. If we denote
\begin{equation}
\frac{\lambda}{\mu} = \alpha \label{NZW-7}
\end{equation}
we can re-write Eq. (\ref{NZW-6}) as
\begin{eqnarray}
&& - \frac{\alpha}{(N+1)^a}P(N) - N^b P(N) + \nonumber\\
&&\hspace{8mm} + \frac{\alpha}{N^a}P(N-1) + (N+1)^b P(N+1) = 0\label{NZW-8}
\end{eqnarray}
which leads to the recurrent relation
\begin{equation}
(N+1)^bP(N+1) = \frac{\alpha}{(N+1)^a}P(N). \label{NZW-9}
\end{equation}
Further simplifications can be made by writing
\begin{equation}
a+b = \nu \label{Add-1}
\end{equation}
which gives us
\begin{equation}
\frac{P(N+1)}{P(N)} = \frac{\alpha}{(N+1)^{\nu}}. \label{NZW-10}
\end{equation}
For $\delta = \nu -1$, this is just the recurrent form of COM-Poisson distribution defined by Eq. (\ref{ZW-4}).  The condition in Eq. (\ref{Add-1}) has allowed us to successfully re-parameterize the recurrent relation of Eq. (\ref{NZW-10}) using $\alpha$ and $\nu=a+b$. This corresponds to the entire class of Markov processes previously characterized by parameters $a$ and $b$ with the birth and death rates given in Eq. (\ref{NZW-5}).

However, to describe the distribution defined by equations (\ref{NZW-2}) and (\ref{NZW-3}) (which is no longer COM-Poisson distribution) still using the birth-death process we need to add an additional term $\lambda^{\star}$ to the birth rate $\lambda_N$ in the form given by
\begin{equation}
\lambda_N = \frac{\lambda}{(N+1)^a} + \lambda^{\star}(N+1)^{(b-1)}. \label{NZW-11}
\end{equation}
If we substitute (\ref{NZW-11}) into (\ref{NZW-4b}) and denote
\begin{equation}
\frac{\lambda^{\star}}{\mu} = \alpha_0\label{NZW-12}
\end{equation}
we get the recurrent relation
\begin{equation}
\frac{P(N+1)}{P(N)} = \frac{\alpha}{(N+1)^{\nu}} + \frac{\alpha_0}{N+1} \label{NZW-13}
\end{equation}
corresponding to Eq. (\ref{NZW-2}).

\section{Summary of results}
\label{sum-res}

The ATLAS data suitable for our purposes cover $5$ different ranges of transverse momentum, each further divided into $4$ different rapidity ranges totalling $20$ different fragments of the phase space. In previous sections we have plotted examples of multiplicity distributions and their corresponding combinants. Figs. \ref{SPD-1} and \ref{SPD-2} show the plots for $4\text{ GeV}<p^{jet}_T<6\text{ GeV}$ at $|y|<1.9$ with $R=0.4$ while Figs. \ref{SPD-5} and \ref{SPD-6} show that for $10\text{ GeV}<p^{jet}_T<15\text{ GeV}$. Instead of showing all figures for $P(N)$ and $C_j$ which we have calculated for each possible $g(N)$ given by Eq. (\ref{NZW-2}) and $P(N)$ using Eq. (\ref{NZW-3}), the presented values of the parameters in are described by the formula
\begin{equation}
\langle N\rangle \left| C_j\right| = A B^j. \label{Res-1}
\end{equation}

With increasing values of transverse momenta of jets, $p_T^{jet}$, the mean multiplicity in jet grows and affects parameters given in the tables. In Table \ref{Table-1} the parameters depend on $\langle N\rangle$ in the following ways:
\begin{eqnarray}
\delta &=& 0.94 + \left(\frac{\langle N\rangle}{4.9}\right)^{4.4}, \label{Res-2}\\
\alpha_0 &=& - 4.94 + 1.7 \langle N\rangle, \label{Res-3}\\
\alpha &\cong& A \cong B \cong 10 + \left(\frac{\langle N\rangle}{3.05} \right)^{6}, \label{Res-4}\\
P(0) &=& 0.5 \exp\left(- \frac{\langle N\rangle}{0.7}\right). \label{Res-5}
\end{eqnarray}

Table \ref{Table-2} shows results for broader jets with $R=0.6$.  Dependence of the parameters on $\langle N\rangle$ in this case are as follows:
\begin{eqnarray}
\delta &=& 0.66 + \left(\frac{\langle N\rangle}{6.38}\right)^{5.85}, \label{Res-6}\\
\alpha_0 &=& - 9.56 + 2.17 \langle N\rangle, \label{Res-7}\\
\alpha &\cong& A \cong B \cong 15.5 + \left(\frac{\langle N\rangle}{5.9} \right)^{17.5}, \label{Res-8}
\end{eqnarray}
Dependence of $P(0)$ on $\langle N\rangle$ remains the same as in Eq. (\ref{Res-5}).

\begin{table*}[h]
\caption {Dependence on $p_T^{jet}$ for $R=0.4$ and $0 < |y| < 1.9$.}
\vspace*{0.2cm}
\begin{center}
\begin{tabular}{|c|c|c|c|c|c|c|c|c|c|}
\hline
             &                    &             &            &           &              &               &               &                \\
 $p^{jet}_T$ [GeV] & $\langle N\rangle$ & $Var(N)$    & ~~ $A$ ~~  &~~ $B$ ~~  &~~ $\alpha$ ~~&~~ $\delta$ ~~ &~~ $\alpha_0$ ~&~~ $P(0)$ ~~    \\
             &                    &             &            &           &              &               &               &                \\
\hline
             &                    &             &            &           &              &               &               &                \\
 $4-6$       &  $3.303$           &  $1.730$    &  $13.3$    & $10.3$    &  $14.596$    &  $1.176$      & $0.286$       &  $0.00430246$  \\
             &                    &             &            &           &              &               &               &                \\
 $6-10$      &  $4.131$           &  $2.585$    &  $15.7$    & $13.0$    &  $16.287$    &  $1.387$      & $2.096$       &  $0.00155689$  \\
             &                    &             &            &           &              &               &               &                \\
 $10-15$     &  $5.170$           &  $3.946$    &  $32.7$    & $27.9$    &  $29.374$    &  $2.220$      & $4.367$       &  $0.000320975$ \\
             &                    &             &            &           &              &               &               &                \\
 $15-24$     &  $6.254$           &  $5.612$    &  $127.0$   & $125.0$   & $126.666$    &  $3.938$      & $6.055$       & $0.0000309069$ \\
             &                    &             &            &           &              &               &               &                \\
 $24-40$     &  $7.644$           &  $7.584$    &  $196.0$   & $180.0$   & $179.212$    &  $7.0$        & $7.63$        & $0.0000160399$ \\
             &                    &             &            &           &              &               &               &                \\
\hline
\end{tabular}
\end{center}
\label{Table-1}
\end{table*}

\begin{table*}[h]
\caption {Dependence on $p_T^{jet}$ for $R=0.6$ and $0 < |y| < 1.9$.}
\vspace*{0.2cm}
\begin{center}
\begin{tabular}{|c|c|c|c|c|c|c|c|c|c|}
\hline
             &                    &             &            &           &              &               &               &                \\
 $p^{jet}_T$ [GeV] & $\langle N\rangle$ & $Var(N)$    & ~~ $A$ ~~  &~~ $B$ ~~  &~~ $\alpha$ ~~&~~ $\delta$ ~~ &~~ $\alpha_0$ ~&~~ $P(0)$ ~~    \\
             &                    &             &            &           &              &               &               &                \\
\hline
             &                    &             &            &           &              &               &               &                \\
 $4-6$       &  $4.398$           &  $2.319$    &  $17.4$    &  $13.2$   &  $20.216$    &  $0.961089$   & $0.0$         &  $0.000767709$ \\
             &                    &             &            &           &              &               &               &                \\
 $6-10$      &  $5.717$           &  $3.878$    &  $15.7$    &  $11.0$   &  $16.709$    &  $0.914$      & $2.528$       &  $0.000272948$ \\
             &                    &             &            &           &              &               &               &                \\
 $10-15$     &  $7.374$           &  $6.878$    &  $63.2$    & $58.0$    &  $58.875$    &  $3.167$      & $7.209$       & $0.0000273138$ \\
             &                    &             &            &           &              &               &               &                \\
 $15-24$     &  $8.526$           &  $8.420$    &  $577.7$   & $580.0$   & $579.390$    &  $6.168$      & $68.509$      &$0.00000141011$ \\
             &                    &             &            &           &              &               &               &                 \\
 $24-40$     &  $9.790$           &  $10.775$   &  $3.3$     & $5.9$     & $226.418$    &  $-0.00412$   & $-218.809$    & $0.000123035$  \\
             &                    &             &            &           &              &               &               &                \\
\hline
\end{tabular}
\end{center}
\label{Table-2}
\end{table*}

\begin{table*}[h]
\caption {Dependence on rapidity interval $|y|$ for $R=0.4$ and $10 < p_T^{jet} < 15$ GeV.}
\vspace*{0.2cm}
\begin{center}
\begin{tabular}{|c|c|c|c|c|c|c|c|c|}
\hline
             &                    &             &           &           &              &               &               &                \\
 $|y|$       & $\langle N\rangle$ & $Var(N)$    &~~ $A$ ~~  &~~ $B$ ~~  &~~ $\alpha$ ~~&~~ $\delta$ ~~ &~~ $\alpha_0$ ~&~~ $P(0)$ ~~    \\
             &                    &             &           &           &              &               &               &                \\
\hline
             &                    &             &           &           &              &               &               &                \\
 $0-0.5$     &  $5.246$           &  $3.086$    & $24.4$    & $27.0$    &  $27.861$    &  $2.227$      & $4.507$       &  $0.000330263$ \\
             &                    &             &           &           &              &               &               &                \\
 $0.5-1$     &  $5.178$           &  $3.949$    & $47.3$    & $44.0$    &  $46.500$    &  $2.631$      & $4.483$       &  $0.000185273$ \\
             &                    &             &           &           &              &               &               &                \\
 $1-1.5$     &  $5.158$           &  $3.852$    & $15.7$    & $17.0$    &  $18.476$    &  $1.677$      & $4.012$       &  $0.00052248$  \\
             &                    &             &           &           &              &               &               &                \\
 $1.5-1.9$   &  $5.061$           &  $3.854$    & $30.7$    & $31.0$    &  $32.632$    &  $2.370$      & $4.310$       & $0.000320779$  \\
             &                    &             &           &           &              &               &               &                \\
\hline
\end{tabular}
\end{center}
\label{Table-3}
\end{table*}

The value of $\langle N\rangle$ increases by a factor of $1.85$ for $R=0.6$ in comparison to the case of $R=0.4$ as seen from the comparison of Tables \ref{Table-1} and \ref{Table-2}. Other dependencies on $\langle N\rangle$ are similar to those from the above equations. However, in the interval of $24$~GeV$ < p_T^{jet} < 40$ GeV, we have $\delta < 0$ and hence, this $p_T^{jet}$ interval was omitted in determining the parameter dependence on $\langle N\rangle$.

With regards to the relation between $Var(N)$ and $\langle N\rangle$, it is observed that for $R=0.4$, we have dispersion $\sigma = \sqrt{Var(N)} = 0.33  \langle N\rangle + 0.23$ and for $R=0.6$ we have $\sqrt{Var(N)} = 0.33 \langle N\rangle + 0.10$~\footnote{It is worth recalling at this point that this linear relationship, $\sqrt{Var(N)} = a\langle N\rangle +b$, known as Wr\'oblewski's law \cite{AKW}, is satisfied for a wide range of multiparticle production processes like $pp$ collisions (here $a = - b = 0.585$), both $\pi^+p$ and $\pi^-p$ collisions (with $a = 0.44$ for $\pi^{\pm}$ and $b = - 0.22$ for $\pi^+$ and $b = -0.9$ for $\pi^-$) \cite{LVH} and for $e^+e^-$ (where $a = 0.25$ and $b = 0.7$)) \cite{SWW}. }.

Note that mean multiplicities $\langle N\rangle$ within the various rapidity intervals do not depend significantly on the rapidity interval $|y|$, as shown in Table~\ref{Table-3}, and are similar to $\langle N\rangle$ for the maximal interval $0 < |y| < 1.9$. Similarly, the other parameters do not differ significantly from the corresponding parameters in the maximum rapidity range (no systematic changes with the width of the rapidity range).

\section{Conclusions}
\label{Concl}

Recurrent relation $g(N)=(N+1)P(N+1)/P(N)$ leads to multiplicity distributions of the form
\begin{equation}
P(N) = \frac{P(0)}{N!}\prod_{i=0}^{N-1} g(i). \label{Rec-1}
\end{equation}
For $g(N)$ given by Eq. (\ref{NZW-2}) with $\alpha_0 = 0$, we have PD for $\delta = 0$, a sub-Poissonian distribution for $\delta > 0$ also known as the Conway-Maxwell-Poisson distribution (COM-PD) \cite{COM1962,Shmuel,ChI,ChShmuel,LZH} and a super-Poissonian distribution for $\delta < 0$.

It turns out, however, that the multiplicity distributions in jets prefer the recurrent relation with $\alpha_0 \neq 0$ leading to multiplicity distributions of the form given by Eq. (\ref{NZW-3}). For small $\langle N\rangle$ we observe sub-Poissonian distributions with modified combinants oscillating as
\begin{equation}
\langle N\rangle C_j \propto (-1)^j \alpha^{j+1}. \label{Rec-2}
\end{equation}
Parameters of multiplicity distributions ($\alpha$, $\alpha_0$ and $\delta$) depend on $\langle N\rangle$. For large $\langle N\rangle$, distributions are super-Poissonian when $-1 < \delta <0$ \footnote{For $\delta = -1$  we have well known NBD, for  $\alpha_0 > \alpha > 0$ (where parameter $k = \alpha_0/\alpha + 1$, if additionally $\alpha_0=0$ we would have geometrical distribution, the most wide one) or BD for $\alpha_0 > 0$ and  $\alpha < 0$ (where parameter $K = - \alpha_0/\alpha - 1$).}.

A sub-Poissonian distribution with $Var(N) < \langle N\rangle$ has not been very popular in the majority of discussions about multiplicity distributions in high-energy physics so far (mainly due the fact that this phenomenon is only observed when $\langle N\rangle$ is not too large).  However, it is quite an important distribution \cite{CCS,PCarruthers}. The existence of a sub-Poissonian distribution
implies at least one of the following scenarios: $(1)$ the underlying elementary processes are not totally random (partially deterministic); $(2)$ the classical  Markov processes describing them require further generalizations. Either conclusion forces us to modify the successful stochastic approach. In our case, we show that the sub-Poisonian multiplicity distributions describing the experimental data can be naturally interpreted as stochastic Markov processes in which the birth and death rates are both multiplicity-dependent.

\vspace*{0.3cm}
\centerline{\bf Acknowledgements}
\vspace*{0.3cm}
This research was supported in part by the National Science Centre, Poland (NCN) Grant 2020/39/O/ST2/00277 (MR) and by grants UMO-2016/22/M/ST2/00176 and DIR/WK/2016/2018/17-1 (GW). H.W. Ang would like to thank the NUS Research Scholarship for supporting this study. In preparation of this work we used  the resources of the Center for Computation and Computational Modeling of the Faculty of Exact and Natural Sciences of the Jan Kochanowski University in Kielce.

\appendix

\section{Relationship of combinants with factorial moments and cumulants}
\label{FqKq}

Usually information contained in $P(N)$ is obtained by examining their corresponding factorial moments, $F_q$, and cumulant factorial moments, $K_q$, (or their ratios) (cf., \cite{Kittel,Book-BP}),
\begin{equation}
K_q = F_q - \sum_{i=1}^{q-1}\binom{q-1}{i-1} K_{q-i}F_i, \label{cumfactmom}
\end{equation}
where
\begin{equation}
F_q = \sum_{N=q}^{\infty} N(N-1)(N-2)\dots(N-q+1)P(N), \label{factmom}
\end{equation}
are the factorial moments. As shown in \cite{Ours2,Ours-EPJA} the $K_q$ can be expressed as an infinite series of the $C_j$,
\begin{equation}
K_q = \sum_{j=q}^{\infty}\frac{(j-1)!}{(j-q)!}\langle N\rangle C_{j-1}, \label{KconC}
\end{equation}
and, conversely, the $C_j$ can be expressed in terms of the $K_q$ \cite{Kittel,Book-BP},
\begin{equation}
C_j = \frac{1}{\langle N\rangle} \frac{1}{(j-1)!} \sum_{p=0}^{\infty}\frac{(-1)^p}{p!}K_{p+j}. \label{ConK}
\end{equation}
Note that $C_j$ depends only on multiplicities smaller than their rank \cite{Combinants-1,Combinants-2} while moments  $K_q$ require the knowledge of all $P(N)$ and therefore are very sensitive to possible limitations of the available phase space \cite{Kittel,Book-BP}. On the other hand,  calculations of combinants require the knowledge of $P(0)$ which may not always be available. Both $C_j$ and $K_q$ exhibit the property of additivity.

\section{Multiplicities from two Binomial Distributions}
\label{2BD}

If we have two sources producing $ N_1 $ and $ N_2 $ particles respectively and each distributed according to BD defined by parameters $\left(K_1,p_1\right)$ and $\left(K_2,p_2\right)$, then the distribution of $N= N_1 + N_2$ particles,
\begin{equation}
P(N) = \sum_{i=0}^{min\left( N,K_1,K_2\right)} P_1(i) P_2(N-i), \label{AZ-1A}
\end{equation}
is described by a generating function comprising the product of generating functions for both sources, i.e. by
\begin{equation}
G(z) = \left( 1 - p_1 + p_1 z\right)^{K_1}\cdot \left( 1 - p_2 + p_2 z\right)^{K_2} \label{AZ-1}
\end{equation}
In this case, the first two moments of the distribution $P(N)$ are given by
\begin{figure}[t]
\begin{center}
\includegraphics[scale=0.48]{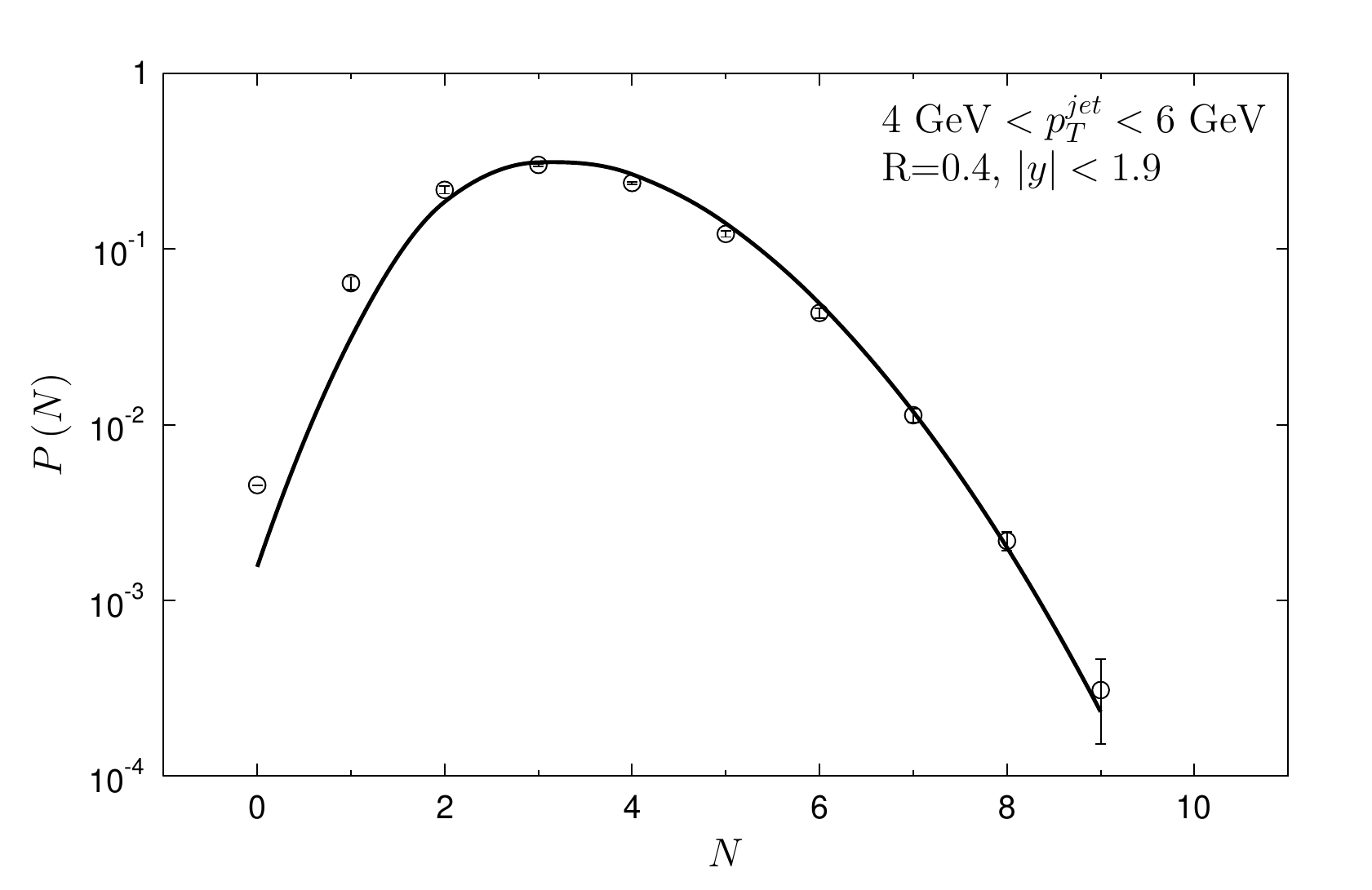}
\end{center}
\vspace{-3mm}
\caption{Points: $P(N)$ from ATLAS data for jets with $4$ GeV $< p^{jet}_T < 6$ GeV and radius parameter $R = 0.4$, over the full measured rapidity range $|y| < 1.9$. The curve fitting these data comes from generating function given by Eq. (\ref{AZ-1}).}
\label{SPD-7}
\end{figure}
\begin{figure}[b]
\begin{center}
\includegraphics[scale=0.48]{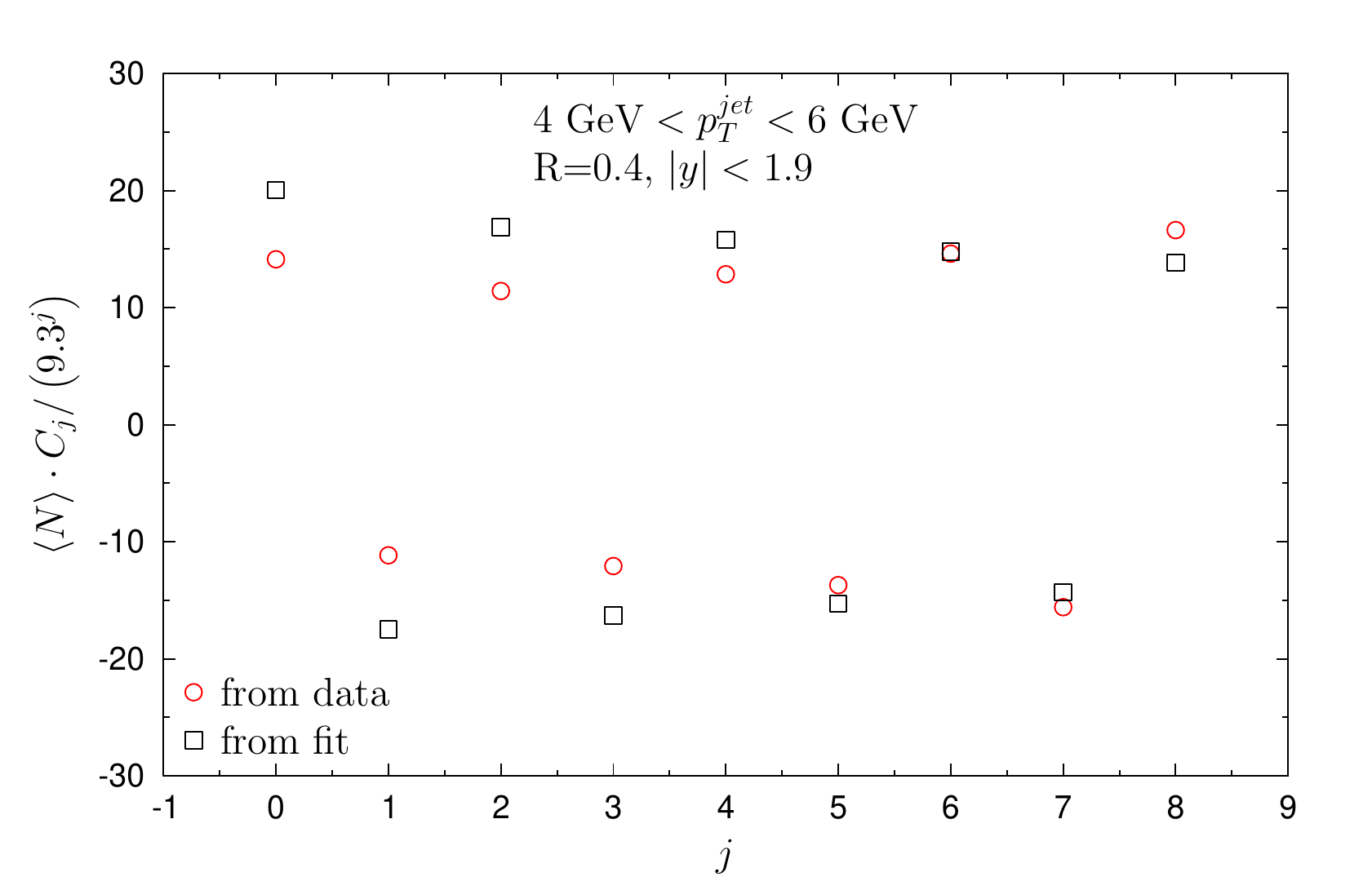}
\end{center}
\vspace{-3mm}
\caption{Comparison of $C_j$ for multiplicity distributions from Fig. \ref{SPD-7}. Red circles: $C_j$ from data on $P(N)$ , black squares: $C_j$ from Eq. (\ref{AZ-3} obtained from theoretical $P(N)$ defined by the generating function given by Eq. (\ref{AZ-1}).}
\label{SPD-8}
\end{figure}
\begin{eqnarray}
\langle N\rangle &=& \frac{d G(z)}{dz}\bigg|_{z=1}  = K_1 p_1 + K_2 p_2, \label{AZ-2}\\
Var(N)&=& \frac{d^2 G(z)}{d z^2}\bigg|_{z=1} + \langle N\rangle - \langle N\rangle ^2 = \nonumber\\
&=& K_1 p_1 (1 - p_1) + K_2 p_2 (1 - p_2). \label{AZ-3}
\end{eqnarray}

Denoting the modified combinant of the first and second BD component as $C(1)_j$ and $C(2)_j$ respectively, the overall modified combinant $\langle N\rangle C_j$ can be written as
\begin{eqnarray}
\langle N\rangle C_j &=& \langle N_1\rangle C(1)_j + \langle N_2\rangle C(2)_j = \label{AZ-4}\\
&=& (-1)^j\left\{ K_1\!\left[ \frac{p_1}{\left( 1 - p_1 \right)}\right]^{j+1}\! +\! K_2\! \left[\frac{p_2}{\left( 1 - p_2\right)}\right]^{j+1}\right\}.\nonumber
\end{eqnarray}

The value of one of the $p$ parameters must be carefully chosen to reflect the observed increase in the amplitude of $C_j$. Choosing (indicative) $K_1 = 2$, $p_1 = 0.9$ and $K_2 = 10$, $p_2 = 0.17$ we have $P(N)$ and $C_j$ as shown in Figs. \ref{SPD-7} and \ref{SPD-8}. For this set of parameters used to fit data, we have $\langle N\rangle = 3.5$ and $Var(N)= 1.6$.

While it is possible to reasonably describe either $P(N)$ or $C_j$ with a suitable choice of parameters, it is not yet possible to describe both simultaneously. Figs. \ref{SPD-7} and \ref{SPD-8} show the extent of deviation from data of such an approach.

It turns out that while by suitable choice of parameters, we can describe (more or less reasonably) separately $P(N)$ or $C_j$, but not simultaneously both observables. Figs. \ref{SPD-7} and \ref{SPD-8} demonstrate to what extent is it possible to get closer to this goal in such an approach.

\end{document}